\documentclass[12pt]{article}
\usepackage{jheppub}

\usepackage{amsmath,graphicx,amssymb,bbm,subfigure}
\usepackage[all]{xy}

\def\be{\begin{equation}}
\def\ee{\end{equation}}
\def\ba{\begin{eqnarray}}
\def\ea{\end{eqnarray}}

\def\r{\rho}
\def\a{\alpha}

\def\g{\gamma}

\def\D{\Delta}

\def\th{\theta}

\def\m{\mu}
\def\n{\nu}
\def\om{\omega}
\def\Om{\Omega}
\def\l{\lambda}
\def\L{\Lambda}
\def\s{\sigma}

\def\cG{{\cal G}}
\def\cH{{\cal H}}
\def\cJ{{\cal J}}
\def\cL{{\cal L}}

\def\cM{{\cal M}}
\def\cN{{\cal N}}
\def\cO{{\cal O}}

\def\no{\noindent}

\def\qq{\qquad}

\def\IR{\relax{\rm I\kern-.18em R}}

\def \ov {\over}

\title{Fermionic impurities in the unquenched ABJM}

\author{Georgios Itsios$^{1,3}$,}
\author{Konstadinos Sfetsos$^{2,1}$,}
\author{Dimitrios Zoakos$^3$}

\affiliation{$^1$ Department of Engineering Sciences, University of Patras,
26110 Patras, Greece}
\affiliation{$^2$  Department of Mathematics, University of Surrey, Guildford GU2 7XH, UK}
\affiliation{$^3$ Centro de F\'\i sica do Porto \& Departamento de F\'\i sica e Astronomia,
Faculdade de Ci\^encias da Universidade do Porto,
Rua do Campo Alegre 687, 4169--007 Porto, Portugal}

\emailAdd{gitsios@upatras.gr}
\emailAdd{k.sfetsos@surrey.ac.uk}
\emailAdd{dimitrios.zoakos@fc.up.pt}

\abstract{We study, in a holographic setup, the effect of adding localized fermionic impurities to the three-dimensional
Chern--Simons-matter theories with unquenched fields in the fundamental representation
of the gauge group. The impurities are introduced as probe D6-branes
extending along the radial direction and wrapping a five-dimensional submanifold
inside a squashed $\mathbb{CP}^3$. We analyze the straight flux tube embeddings and study the corresponding
fluctuation modes of the D6-branes. The conformal dimensions of the
operators dual to such fluctuations depend non-trivially on the ratio of the flavor number to the Chern--Simon level of
the unquenched ABJM.}

\keywords{Gauge-gravity correspondence}

\subheader{DMUS-MP-12/07}

\begin{document}

\def\baselinestretch{1.2}
\baselineskip 20 pt
\no

\maketitle


\section{Introduction}

One of the developments emerging from string theory explorations is the idea of a
gauge/gravity correspondence \cite{Maldacena:1997re}.
The remarkable feature of the correspondence is the relation of the
strongly coupled regime of the gauge theory to the weakly coupled regime of the string theory and
vice--versa. Consequently, it has become a powerful tool in studying strongly interacting
systems, allowing for novel computations that go beyond the standard perturbative techniques of
quantum field theories.

\paragraph{}

In this theme, the recent developments on the $AdS_4/CFT_3$ constitute a
rich framework in which fundamental questions about the correspondence
can be posted. In particular, the ABJM theory \cite{Aharony:2008ug} is a $U(N)\times U(N)$ Chern--Simons gauge theory
with levels $(k,-k)$ and bifundamental matter fields. In the large $N$ limit the theory acquires
a supergravity description in terms of the $AdS_4\times {\mathbb S}^7/{\mathbb Z}_k$ geometry. When the
Chern--Simons level is large the size of the fiber is small and the system acquires a
ten-dimensional description in terms of $AdS_4\times {\mathbb C}{\mathbb P}^3$
with fluxes, that preserves 24 supersymmetries.
The addition of flavor to the ABJM theory
(fields transforming in the fundamental representations
$(N,1)$ and $(1,N)$ of the $U(N)\times U(N)$ gauge group) is realized through D6-branes filling the
$AdS_4$ space and wrapping a submanifold inside the ${\mathbb C}{\mathbb P}^3$, while preserving
a fraction of the initial supersymmetry \cite{Hohenegger:2009as}.  The addition of a
large number of flavor branes, continuously smeared in the transverse space, produces a backreaction
of the original geometry and induces a deformation.
Utilizing techniques developed in \cite{Bigazzi:2005md,Casero:2006pt} and reviewed in \cite{Nunez:2010sf},
this unquenched solution was computed in \cite{Conde:2011sw} and depends non-trivially
on the number of flavors.

\paragraph{}

The addition of an extra set of branes interacting with the colored ones creates a defect in the gauge
theory. The characteristic example of this class is ${\cal N}=4$ Super-Yang-Mills (SYM)
with fermionic impurities.
The brane implementation of such a construction is through the addition of D5-branes 
into the $AdS_5\times S^5$ background.  The D5-branes extend along the radial direction and 
wrap  a four-dimensional submanifold inside the five-dimensional sphere, with the corresponding
worldvolume being $AdS_2\times {\mathbb S}^4$ \cite{Camino:2001at}
(for other embeddings with lower dimensional spheres see \cite{Karaiskos:2011kf}).
These configurations have been used recently in \cite{Kachru:2009xf, Benincasa:2011zu}
to holographically construct dimer models, through D-branes that connect impurities on the 
boundary of $AdS$.
The holographic setup of D3- and D5-branes also realizes the maximally supersymmetric Kondo model
\cite{Mueck:2010ja,Harrison:2011fs,Benincasa:2012wu} (see also \cite{Faraggi:2011bb,Faraggi:2011ge}).

 \paragraph{}

In this paper we consider D6-branes extending along an $AdS_2\subset AdS_4$ and a
wrapping of a five-dimensional submanifold inside the squashed ${\mathbb C}{\mathbb P}^3$,
in order to construct the holographic dual of a Chern--Simons-matter theory with flavor and
fermionic impurities. We will build our solution on the unquenched background solution \cite{Conde:2011sw} 
(for other type of related impurities using D8-branes see also \cite{Ammon:2009wc}).

\paragraph{}

An overview of the paper is as follows:
In section \ref{sec:ABJM-Intro} we give a short, self contained, review of the gravity dual
of a three-dimensional Chern--Simons-matter theory with unquenched fields in the fundamental representation of the gauge group.
In section \ref{density} we introduce impurities as probe D6-branes wrapping a five-dimensional
submanifold inside the squashed $\mathbb{CP}^3$ and analyze straight flux tube embeddings.
In section \ref{sec:Fluct} we analyze in detail the fluctuation modes of the D6-branes around the
straight flux tube configurations and compute the conformal dimensions of the operators dual to
such fluctuations. Due to the presence of the unquenched ABJM there is an explicit dependence
on the number of flavors.
In section \ref{sec:Conclusions} we conclude and discuss lines of possible future related research.
In the appendix \ref{ap:apA} we analytically derive the Lagrangian for the fluctuations of the probe brane and
in appendix \ref{ap:apB} the explicit solution of the radial type differential eq. arising in these fluctuations. In
appendix \ref{ap:apC} we compute the spectrum of the Laplacian corresponding to the angular part of the
operator entering into the fluctuation analysis.


\section{Review of the ABJM with unquenched massless flavor} \label{sec:ABJM-Intro}

In this section, following \cite{Conde:2011sw}, we will provide a self contained review of type-IIA supergravity solutions, 
dual to three-dimensional Chern--Simons-matter theories, after the addition of unquenched flavor. 
This is implemented through D6-branes that extend along the Minkowski directions and smear in the internal space,
in a way that preserves ${\cal N}=1$ supersymmetry.  The geometry
is $AdS_4\times {\cal M}_6$, where ${\cal M}_6$ is the squashed Fubini--Study metric of  ${\mathbb C}{\mathbb P}^3$
\cite{Nilsson:1984bj} and the squashing factors depend on the number of flavors.
The metric of the flavored ABJM background (in the string frame) is given by
\begin{equation}
ds^{2} \, = \, L^{2} \, ds^{2}_{AdS_{4}}  \, + \,  ds_6^2 \ ,
\label{flamet}
\end{equation}
with the standard parametrization for the $AdS_{4}$ metric\footnote{We have rescaled the Minkowski coordinates 
as $x^{\mu} \rightarrow L^2 x^{\mu}$, while working in units $\alpha'=1$.}
\begin{equation}
ds^{2}_{AdS_{4}} \, = \,  r^{2} \, \left(-dt^{2} \, + \, dx^{2} \, + \, dy^{2} \right) \, + \,  {dr^{2} \ov r^{2}} \, ,
\end{equation}
while the six-dimensional metric is written in terms of the $SU(2)$ instanton of $S^4$
\begin{equation} \label{6dv1}
ds^2_{6}\,=\,{L^2\over b^2}\,\,\Big[\,
q\,ds^2_{{\mathbb S}^4}\,+\,\big(d x^i\,+\, \epsilon^{ijk}\,A_j\,x^k\,\big)^2\,\Big] \, ,
\end{equation}
where $b$ and $q$ are constant squashing factors. The metric for the unit $S^4$ is denoted by $ds^2_{{\mathbb S}^4}$
and $x^i$ ($i=1,2,3$) are Cartesian coordinates parametrizing the unit $S^2$ whereas $A^i$, $i=1,2,3$
are the components of the non-Abelian one-form connection corresponding to the $SU(2)$ instanton.
The solution depends on two integers $N$ and $k$ which, on the gauge theory side, represent
the rank of the gauge group and the Chern--Simons level, respectively. In string units,
the $AdS_4$ radius $L$ can be written as
\begin{equation}
L^4\,=\,2\pi^2\,{N\over  k }\,
{(2-q)\,b^4\over
q(q+\eta q\,-\,\eta)} \, .
\label{AdSradius}
\end{equation}
Introducing a set of $SU(2)$ left-invariant one-forms, which satisfy the usual relation
$d\omega_i={1\over2}\,\epsilon_{ijk}\,\omega_j\wedge\omega_k$, together with the coordinate $\alpha$
($0\leqslant \alpha \leqslant \pi$) one parametrizes the unit $S^4$ as
\begin{equation}
ds^{2}_{S^{4}} \, = \, d\a^{2} \, + \, {\sin^{2}\a \ov 4} \,
\left( \, \om_{1}^{2} \, + \, \om_{2}^{2} \, + \, \om_{3}^{2} \, \right) \, .
\end{equation}
The parameterization for the $\om$'s is
\begin{eqnarray}
 && \om_{1} \, = \, \cos\psi \, d\th_{2} \, + \,  \sin\psi \, \sin\th_{2} \, d\phi_{2}
\nonumber\\
 && \om_{2} \, = \, \sin\psi \, d\th_{2} \, - \,  \cos\psi \, \sin\th_{2} \, d\phi_{2}
 \\
 && \om_{3} \, = \, d\psi \, + \, \cos\th_{2} \, d\phi_{2}\ ,
 \nonumber
\end{eqnarray}
while the $SU(2)$ instanton one-forms $A_i$ are given by
\begin{equation}
A_{i}\,=\,- \, \sin^2 {\alpha \over 2} \, \omega_i \, .
\end{equation}
Parametrizing the $x^i$ coordinates of the ${\mathbb S}^2$ by means of the angles
$\theta_1$ and $\phi_1$, the following relation holds\footnote{Explicitly we have
$x^1=\sin\theta_1\,\cos\phi_1$,
$x^2=\sin\theta_1\,\sin\phi_1$,
$x^3=\cos\theta_1$
with $0\leqslant \theta_1<\pi$ and $0\leqslant \phi_1<2\pi$. For completeness we also note that
$0\leqslant \th_2 <  \pi$, $0\leqslant \phi_2 < 2\pi$ and $0\leqslant \psi <  4\pi$.}
\begin{equation}
\big(d x^i\,+\, \epsilon^{ijk}\,A_j\,A_k\,\big)^2\,=\,E_1^2\,+\,E_2^2\,\,,
\end{equation}
where  $E_1$ and $E_2$ are the following one-forms
\begin{eqnarray}
 && E_{1} \, = \, d\th_1 \, + \, \sin^{2} {\a \ov 2}\left(\om_{1}\sin\phi_1 \, - \, \om_{2}\cos\phi_1 \right)
\nonumber\\
 && E_{2} \, = \, \sin\th_1 \left(d\phi_1 \, - \, \om_{3}\sin^{2} {\a \ov 2}\right) \, + \,
 \sin^{2} {\a \ov 2}\cos\th_1 \left(\om_{1}\cos\phi_1 \, + \, \om_{2}\sin\phi_1 \right)\ .
\end{eqnarray}
Putting all these ingredients together we rewrite the six-dimensional metric \eqref{6dv1} as
\begin{equation} \label{6dv2}
ds^2_6\,=\,
{L^2\over b^2}\,\Big[\,q\,ds^2_{{\mathbb S}^4}\,+\,
E_1^2\,+\,E_2^2\,\Big] \, .
\end{equation}
In order to write the expression for the $F_2$ we will introduce a new set of one-forms
\begin{eqnarray}
&&
S_1\, = \, \sin\phi_1\,\omega_1 \, - \, \cos\phi_1\,\omega_2 \,,
\nonumber \\
&&
S_2 \, = \, \sin\theta_1\,\omega_3 \, - \, \cos\theta_1 \left(\cos\phi_1\,\omega_1 \, + \,
\sin\phi_1\,\omega_2\right)\,,
 \\
&&
S_3 \, = \, - \, \cos\theta_1\,\omega_3 \, - \, \sin\theta_1 \left( \cos\phi_1\,\omega_1\, + \,
\sin\phi_1\,\omega_2\right)\,.
\nonumber
\end{eqnarray}
Then, the ansatz for the  $F_2$ is the following
\begin{equation} \label{F2}
F_2\,=\,{k\over 2}\,\,\Big[\,\,
E_1\wedge E_2\,-\,\eta\,
\big({\cal S}_{\a}\wedge {\cal S}_{3}\,+\,{\cal S}_1\wedge {\cal S}_{2}\big)
\,\,\Big]\,\,,
\end{equation}
where the one-forms ${\cal S}_{\a}$ and ${\cal S}_{i}$ are
\begin{equation} \label{calS}
{\cal S}_{\a}\,=d\a \ ,  \qq
{\cal S}_{i}\,=\,{\sin\a\over 2}\,S_i  \ ,\qq i=1,2,3
\end{equation}
and $\eta$ is a squashing parameter, directly related to the number $N_f$ of flavors as
\begin{equation}
\eta\,=\,1\,+\,{3N_f\over 4k}\ ,\qq 1\leqslant \eta <\infty\ .
\label{eta}
\end{equation}
The internal squashing $q$ is related to $\eta$ through the relation
\begin{equation} \label{q-eta}
q\,=\,{3(1+\eta) - \sqrt{9\eta^2-2\eta+9}\over 2} \ ,
\end{equation}
while $b$ can be written in terms of $q$ and $\eta$ as
\begin{equation} \label{b}
b\,=\,{q(\eta+q)\over 2(q+\eta q-\eta)}\ .
\end{equation}
The dilaton and the $F_4$ have the following expressions
\begin{equation}
e^{-\Phi}\,=\,{b\over 4}\,{\eta+q\over 2-q}\,{k\over L} \ ,\qq
F_4\,=\,{3k\over 4}\,\,\,{(\eta+q)b\over 2-q}\,\,L^2\,\,\Omega_{AdS_4}\ .
\label{dilaton+F4}
\end{equation}
The value of $q$ ranges from $1$ to $5/3$, while the case $q=1$ (no flavors) corresponds to the $\cN=6$ ABJM background. We also note that
had we taken the positive square root in \eqref{q-eta} it would have corresponded to a different branch for which $q\geqslant 5$.
For the value $q=5$ the background has reduced supersymmetry, whose metric is the sum of the $AdS_{4}$ space with the 
squashed $\mathbb{CP}^{3}$. This metric is not the one corresponding to an Einstein space \cite{Awada:1982pk, Ooguri:2008dk}.
This would have corresponded to the value $q=2$, which nevertheless is not allowed in either branch.


\section{The Hamiltonian density of D6-brane probes} \label{density}

In this section we will consider a D6-brane probe extending along the radial direction $r$
and wrapping a five-dimensional submanifold inside the squashed $\mathbb{CP}^3$
at constant values of the spatial Minkowski  directions $x$ and $y$. The background
coordinates $X^{M}$ and the worldvolume coordinates $\zeta^{\m}$ are
\begin{equation}
\begin{array}{ll}
\textrm{\emph{{\rm Background}}:} &
X^{M}\, = \, (\, t, \, x, \, y, \, r, \, \alpha,  \, \theta_2, \, \phi_2,  \, \psi,  \, \theta_1, \, \phi_1 \,)\ ,\\[1em]
\textrm{\emph{\rm Brane}:}      &
\zeta^{\m} \, = \, (\, t, \, r, \, \g^{i} \, ) \, = \, (\, t, \, r, \, \theta_2, \, \phi_2,  \, \psi, \, \theta_1, \, \phi_1 \, )\ .
\end{array}
\label{kjdkh}
\end{equation}
We consider embeddings in which the angle $\a$ depends only on the radial direction, i.e.
\begin{equation}
\a \, = \, \a(r) \, .
\end{equation}
These embeddings correspond to configurations in which the flux tube starts from the boundary
of the  $AdS_4$ and reaches the origin of the holographic coordinate.\footnote{That excludes
hanging flux tubes, namely configurations that reach a minimal value of $r$ and return to the boundary.
In such cases a non-constant Cartesian embedding coordinate is needed.
}
We will also turn on an electric worldvolume gauge field component $F_{0r}$, whose source is the
RR potential $C_5$ through the Wess--Zumino (WZ) term of the D6-brane action.

The Dirac-Born-Infeld (DBI) part of the D6-brane action is given by
\begin{equation}
{\cal S}_{DBI}=-T_{6} \int d^{7}\zeta \; e^{-\Phi} \sqrt{-\textrm{det}(g+F)}\ ,
\end{equation}
where $g$ is the induced metric on the worldvolume of the D6-brane and $T_6$ is the brane tension.
After integrating over all the angles of the internal space we arrive at the following expression for the
DBI contribution to the action
\begin{equation}
S_{DBI}\,=\,\int dt \, dr \,{\cal L}_{DBI} \, ,
\end{equation}
with
\begin{equation} \label{DBI}
{\cal L}_{DBI} \, = \, - \, \frac{N L^2}{8 \pi} \, \frac{b}{\sqrt{q}}
\, \sin^{3}\a \,
\sqrt{\,1 \, + \,  {q \ov b^{2}} \, r^{2} \a^{'2}\, - \, L^{-4}F_{0r}^{2}} \, ,
\end{equation}
where $\alpha'$ denotes $d\alpha/dr$.
The WZ part of the action is given by
\begin{equation}
\cL_{WZ} \, = \, T_{6} \; \int C_{5} \wedge F \, \equiv \, \int dt \, dr \,{\cal L}_{WZ} \, .
\end{equation}
The definition for the RR six-form is through the Hodge dual of the RR four-form, 
namely $\star F_{4} = -F_{6}$, and in this way it is possible to obtain an expression for the
five-form potential $C_5$
\begin{equation}
C_{5} \, = \, - \, {\pi^2 \ov 8} \, N \, C(\a) \, \sin\th_1 \, \sin\th_2 \, d\th_1 \wedge d\th_2 \wedge d\phi_1 \wedge d\phi_2 \wedge d\psi\ ,
\end{equation}
where
\begin{equation}
C(\a) \, = \, \cos\a \, (\, \sin^{2}\a \, + \,  2 \, ) \, - \, 2 \, .
\end{equation}
After integrating over the above five angles $\g^{i}$ we obtain
\begin{equation} \label{WZ}
\cL_{WZ} \, = \, - \, {N \ov 8\pi}  \, C(\a) \, F_{0r} \, .
\end{equation}
The Lagrangian is the sum of the \eqref{DBI} and \eqref{WZ}
\begin{equation} \label{DBI+WZ}
\cL \, = \, - \, {N \ov 8\pi} \, \frac{b}{\sqrt{q}}
\Bigg[ \, L^{2} \, \sin^{3}\a \, \sqrt{1\, + \,{q \ov b^{2}} \, r^{2} \, \a^{'2} \, - \, L^{-4}F_{0r}^{2}} \, +
\, {\sqrt{q} \ov b} \, C(\a) \, F_{0r} \Bigg] \, .
\end{equation}
The equation of motion for the gauge field implies the following
\begin{equation}
{\partial {\cal L}\over \partial F_{0r}}\,=\,{\rm constant}\, .
\end{equation}
This constant is related to the number $n$ of strings (quarks) of the flux tube, through
the quantization condition of \cite{Camino:2001at}
\begin{equation}
{\partial {\cal L}\over \partial F_{0r}}\,=\,n\,T_f\, ,
\label{quantization}
\end{equation}
where  $T_f$ is the tension and  $n\in{\mathbb Z}$ is the charge of the 
fundamental string attached to the D6-brane. Exploiting the above quantization condition,
we obtain
\begin{equation} \label{useful-condition}
 {\sin^{3}\a \ov \sqrt{1\, - \, L^{-4}F_{0r}^{2} \, + \, {q \ov b^{2}} \, r^{2} \, \a^{'2}}} \, = \,
 {\sqrt{\sin^{6}\a \, + \, C_{n}(\a)^{2}} \ov \sqrt{1\, + \, {q \ov b^{2}} \, r^{2} \, \a^{'2}}} \, ,
\end{equation}
where we have defined
\begin{equation} \label{def-Cn}
C_{n}(\a) \, \equiv \, {\sqrt{q} \ov b} \, \left( \, C(\a) \, + \, {4n \ov N} \, \right) \, .
\end{equation}
Then from \eqref{useful-condition} we obtain for the field strength
\begin{equation} \label{F_0r}
F_{0r} \, = \, {L^{2} \, \sqrt{ 1\, + \,  {q \ov b^{2}} \, r^{2} \, \a^{'2} }
\over \sqrt{ \sin^{6}\a \, + \, C_{n}(\a)^{2} }} \, C_{n}(\a) \, .
\end{equation}
In order to eliminate the electric field from the equations of motion
we compute the Hamiltonian of the system by performing a Legendre transformation
in \eqref{DBI+WZ}
\begin{equation}
\cH \, = \, F_{0r} \,  {\partial \cL \ov \partial F_{0r}} \, - \, \cL \, .
\end{equation}
Using the above results together with \eqref{useful-condition} we end up with
the following formula for the Hamiltonian density
\begin{equation} \label{hamilton_density}
\cH \, = \, {N L^{2} \ov 8\pi} \, \frac{b}{\sqrt{q}} \, \sqrt{1\,+ \, {q \ov b^{2}} \, r^{2} \, \a^{'2}} \,
\sqrt{\sin^{6}\a \, + \, C_{n}(\a)^{2}} \, .
\end{equation}
It remains to determine $\a(r)$ by integrating the corresponding Euler-Lagrange equations.
In the next subsection we will constrain our analysis to
configurations with constant $\alpha$. Embeddings with $\alpha$ depending on the holographic
coordinate are related to the baryon vertex of the ABJM theory and will be analyzed in future work
(for the similar analysis in the $AdS_5\times S^5$ case see \cite{Callan:1998iq,Callan:1999zf}).


\subsection{Flux tube configurations} \label{sec:FTC}

In this subsection we will calculate the energy density of the configurations with constant $\a$.
Such configurations must satisfy the condition
\begin{equation}
 \left.{\partial\cH \ov \partial\a}\right|_{\a^{'}=0} \, = \, 0 \, .
\end{equation}
Since
\begin{equation}
 \left.{\partial\cH \ov \partial\a}\right|_{\a^{'}=0}  \, = \,
 3\, {N L^{2} \ov 8\pi} \, \frac{b}{\sqrt{q}} \, {\sin^{3}\a  \, \L_{n}(\a) \ov \sqrt{\sin^{6}\a \,+ \, C_{n}(\a)^{2}}} \, ,
\end{equation}
with
\begin{equation} \label{Lambda}
\L_{n}(\a) \, \equiv \, \sin^{2}\a \, \cos\a \, - \, {\sqrt{q} \ov b} \, C_{n}(\a)  \, ,
\end{equation}
the non-trivial configurations with constant $\a$ are the solutions of the following algebraic equation
\begin{equation}
\L_{n}(\a) \, = \, 0 \, ,
\label{lnna}
\end{equation}
which can be written as a cubic equation in $\cos\a$ as
\begin{equation} \label{cubic}
 \left(\,1 \, - \, {q \ov b^{2}} \, \right)\cos^{3}\a \, - \, \left( \, 1 \, - \, {3q \ov b^{2}} \, \right)\cos\a \, - \,
 {2q \ov b^{2}} \, \left( \, 1 \, - \, {2n \ov N} \, \right) \, = \, 0 \, .
\end{equation}
Due to Bolzano's theorem\footnote{Note that
 ${q / b^{2}} \in \left[1,{16 \ov 15}\right]$, $\Lambda_{n}(\alpha)$ runs monotonically as $\alpha \in [0,\pi]$ and
\begin{equation}
 \L_n(0)= -4 {q\ov b^2} {n\ov N}\leqslant 0  \ ,\qq \L_n(\pi) = 4 {q\ov b^2}\left(1- {n\ov N}\right)\geqslant 0 \ .
 \end{equation}
 } the function $\L_{n}(\a)$ has at least one root in the
interval $\a \in [0,\pi]$ for every $n$ in the range $0\leqslant n\leqslant N$,
while the monotonicity of the function in this interval tells us that the root
is unique. After using \eqref{Lambda} to express $C_{n}(\a_{n})$ in terms of $\a_{n}$
\begin{equation}
C_{n}(\a_n)  \, = \, {b \over \sqrt{q}} \, \sin^{2}\a_n \, \cos\a_n  \, ,
\end{equation}
as well as \eqref{hamilton_density}, we obtain the energy density of the
configurations with constant $\a$
\begin{equation} \label{energy}
E_{n} \, =  \, {N L^{2} \ov 8\pi} \, \frac{b^2}{q} \,  \sin^2\a_{n}
\sqrt{ \cos^2\a_{n} \, + \, {q \over b^2} \, \sin^2\a_{n} } \, ,
\end{equation}
where $\a_{n}$ is a solution of \eqref{cubic}.\footnote{For $n\ll N$ the energy turns out to be simply the
sum of the energies of the individual fundamental strings, i.e. $E_n \simeq {n\ov 2\pi} L^2$, where the extra
factor $L^2$ is related to the overall appearance of the same factor in \eqref{flamet}
due to a rescaling of the world-volume coordinates. It is worth noting that in this dilute type approximation there
is no dependence of the energy on the flavor number.
}
The constant electric field $F_{0r}$ corresponding to such configurations is computed from
\eqref{F_0r}. One finds that
\begin{equation} \label{gauge}
\bar{f}_{0r} \, = \, {L^{2} \, \cos \a_{n} \ov \sqrt{ \cos^2\a_{n} \, + \, {q \over b^2} \, \sin^2\a_{n} }} \, .
\end{equation}
Notice that from \eqref{cubic} we have
\begin{equation} \label{change}
\Lambda_{n} (\alpha_n) \, = \, - \, \Lambda_{N-n} (\pi - \alpha_n) \qq \Longrightarrow \qq
\alpha_{N - n} \, = \, \pi \,  - \,  \alpha_n \, ,
\end{equation}
which combined with \eqref{energy}, is telling us that $E_n$ is invariant
under the change $n\to N-n$, as it should if an
object is to transform in the fully anti-symmetric representation of the gauge group, with $n$ being the number
of boxes in the corresponding Young tableaux.
The induced metric on the D6-brane worldvolume is
\begin{equation}
d s^2\,=\,
L^2\,\,\Big[\,-\,r^2\,dt^2\,+\,{dr^2\over r^2}\,+\,d s^2_{M_5}\,\Big] \, ,
\label{induced-metric}
\end{equation}
which is of the form $AdS_2\times M_5$,
with the line element of $M_5$ having the following expression
\begin{equation} \label{M5}
ds^{2}_{M_{5}} \, = \, \tilde{g}_{ij} \, d\g^{i} \, d\g^{j} \, = \,
{q \ov 4b^{2}} \, \sin^{2}\a_{n}  \, \left( \,  \om_{1}^{2} \, + \,  \om_{2}^{2} \, + \,  \om_{3}^{2} \, \right) \, +
\, {1 \ov b^{2}} \, \left(\,  E_{1}^{2} \, + \, E_{2}^{2} \, \right) \, .
\end{equation}
Under the change of $\a_{n}$ described in \eqref{change} the line element of \eqref{M5}
remains invariant as well.


\section{Fluctuations of the impurities} \label{sec:Fluct}

Moving one step forward, we will study in this section fluctuations around the static configurations we have computed.
For this reason we consider the following
\begin{equation} \label{perturbation}
\a \, = \, \a_{n} \, + \, \xi(\zeta) \, , \quad  F \, = \, \bar{f} \, + \,  f(\zeta) \, , \quad x \, = \, \bar{x} \, + \, \chi(\zeta) \, ,
\end{equation}
where $\a_{n}$ is a solution of the condition $\L_{n}(\a)=0$,
$\bar x$ is the constant Cartesian coordinate of the unperturbed D6-brane
and $\bar{f}$ is the background gauge field strength with non-vanishing component given by \eqref{gauge}.
The fluctuations around these constant values, namely $\xi, \; \chi \;$ and $ \; f $,
depend, as indicated, on the D-brane coordinates $\zeta^{\m}$ in \eqref{kjdkh}.
The total perturbed Lagrangian density is the sum of the DBI and the WZ parts,
and a detailed derivation is presented in the appendix \ref{ap:apA}. Indeed, if in
\eqref{flloicr} we neglect constant and total derivative terms
we find the following Lagrangian density for the quadratic fluctuations
\begin{eqnarray}
\label{2nd-Lang}
\mathcal{L} &=& - \, T_{6} \,  {\pi^{2} \, N \, b^{6} \, L^{2} \ov q^{2}} \; P^{1/2} \, \sqrt{\tilde{g}}
\left\{ {\, 1 \ov 2} \, L^{2} \, r^{2} \, \cG^{\m\n} \, \partial_{\m}\chi \, \partial_{\n}\chi  \, + \,
{1 \ov 2} \, {q \ov b^{2}} \, L^{2} \, \cG^{\m\n}\partial_{\m}\xi \, \partial_{\n}\xi  \right.
\nonumber\\ [5pt]
&& \left. \, + \,  {1 \ov 4} \, \cG^{\m\r}\cG^{\n\s}f_{\m\n}f_{\r\s} \, + \, V \, \xi^{2} \, - \,
W \, \xi \, f_{0r} \,
\right\}\ ,
\end{eqnarray}
where we have for notational convenience set
\begin{eqnarray}
&& P \, = \,  {\sin^{6}\a_{n} \ov \sin^{6}\a_{n} \, + \,  C_{n}(\a_{n})^{2}} \, , \quad
V \, = \, - \, {3 \ov 2 \, \sin^{2}\a_{n}} \, ,
\nonumber \\ [5pt]
&& W \, = \, {3 \ov L^{2} \, P^{1/2} \, \sin^{3}\a_{n}} \, \Big( C_{n}(\a_{n}) \, \cot\a_{n} \, + \,
\frac{\sqrt{q}}{b} \, \sin^{3}\a_{n} \Big) \,
\end{eqnarray}
and the seven-dimensional metric $\cG$ is defined in \eqref{sym}.


\subsection{Fluctuation of the Cartesian coordinate}

Here we study the fluctuations of the Cartesian coordinate which do not couple to those
of the gauge field and of the embedding angular coordinate.
The equation of motion for $\chi$ can be easily derived from \eqref{2nd-Lang} and it is
\begin{equation}
 \partial_{r}(r^{4}\partial_{r}\chi) \, - \, \partial_{0}^{2}\chi \, + \, r^{2} \; P \; \nabla_{M_{5}}^{2}\chi \, = \, 0 \, .
\end{equation}
To obtain the above equation of motion we explicitly used the components of $\cG^{\m\n}$,
while $\nabla_{M_{5}}^{2}$ is the Laplacian operator of the five-dimensional manifold in
\eqref{M5} (see appendix \ref{ap:apC}). The actual expression for this operator is quite
complicated. Remarkably, we were able to find explicit solutions of the form either
$\chi=\chi(t,r,\th_2,\phi_2,\psi)$ or $\chi=\chi(t,r,\th_1,\phi_1)$ (see appendices \ref{ap:apC1} and
\ref{ap:apC2}, respectively). Without loss of generality in the following analysis of the conformal dimensions
we will focus on the first class of solutions.
Using the separation of variables
\begin{equation}
\chi \, = \, e^{\imath Et} \, R(r) \, \Om(\th_2,\phi_2,\psi) \, ,
\end{equation}
and equation \eqref{3Laplacian} from appendix \ref{ap:apC1}
\begin{equation}
 \nabla^{2}_{M_{5}}\Om \, = \, {b^{2} \ov q \, \sin^{2}\alpha_{n}} \, \nabla_{S^{3}} \Om \, = \, -
 \, {b^{2} \ov q} \,  {l(l+2) \ov \sin^{2}\a_{n}} \, \Om, \qquad  l=0,1,2,\ldots
\end{equation}
the equation of motion for the radial function $R(r)$ becomes
\begin{equation} \label{eq:ODE}
\partial_{r}\left( \, r^{4} \, \partial_{r}R \, \right) \, + \,
\left(\,  E^{2} \, - \,  {b^{2} \ov q} \; {l(l+2) \ov \sin^{2}\a_{n}} \; P \; r^{2} \, \right) \, R \, = \, 0 \, .
\end{equation}
This equation can be solved exactly, but since we are interested in
the asymptotic behavior of the solution we put all the details on the analytic derivation
in appendix \ref{ap:apB}.
Assuming that $R(r)  \sim r^{\l}$ at large $r$, we arrive to the following quadratic equation
\begin{equation} \label{chi-quad}
\l(\l+3) \, = \,  {b^{2} \ov q} \; {l(l+2) \ov \sin^{2}\a_{n}} \; P \, ,
\end{equation}
with solutions that we will denote as $\l_1$ and $\l_2$. We would like to associate them
with the dimensions $\D$ of the operators of the defect theory. The fluctuations
are not canonically normalized since there is a factor of $r^2$ in front
of the kinetic term for the field $\chi$ in \eqref{2nd-Lang}.
Hence, we cannot simply use the usual relation \cite{Witten:1998qj}
\be
\D = {d\ov 2} +\sqrt{{d^2\ov 4} + m^2}\ ,
\label{df35}
\ee
with $d=1$.
Instead one may employ an approach that gives the result immediately \cite{Kruczenski:2003be}.
According to this, if a scalar field in $AdS_{2}$ at large $r$ behaves as
\be
\chi \, \sim \, d_{1}\, r^{-2 \l_{1}} \, + \,  d_{2} \, r^{-2\l_{2}}, \quad \l_{2}>\l_{1} \, ,
\ee
then the dimension of the operator dual to the normalizable mode is
\be
\Delta \, = \, {1 \ov 2} \, + \, \l_{2} \, - \, \l_{1} \, .
\ee
In our case the conformal dimension becomes
\begin{equation}
\label{dimension-cartecian}
\D \, = \,  {1 \ov 2} \, + \,  \sqrt{ \, {9 \ov 4}  \, + \,   {b^{2} \ov q} \;
{l(l+2) \ov \sin^{2}\a_{n}} \; P } = {1 \ov 2} \, + \,  \sqrt{ \,  {9 \ov 4} \, + \,   {l(l+2)\ov
q/b^2 \sin^2\a_n + \cos^2\a_n}} \, .
\end{equation}
In general, the conformal dimension is not a rational number and
$\Delta=2$ for $l=0$. Moreover the dimension depends on the filling fraction $\nu$, through the
dependence of the angle $\alpha_n$, and is invariant under the change of \eqref{change}.
Due to the range of the parameter $q/b^2$, defined in footnote 2, the prefactor multiplying $l(l+2)$ is of order
one.\footnote{Another way to arrive at the same result is to redefine
the fluctuations by absorbing the factor $r$ in the kinetic for $\chi$ in \eqref{2nd-Lang} into a new scalar
field $\phi = r\chi$. Then $\phi$ becomes canonically normalized but after some algebraic manipulations one
sees that $m^2$ is shifted by a factor of 2. Then after using \eqref{df35}, with $d=1$, one derives \eqref{dimension-cartecian}.
}
The dependence of the conformal dimension on the number of flavors becomes in general complicated
and it admits a power series expansion around the unquenched result. In particular,
in the half filling fraction case,  $\nu \, = \, \frac{1}{2}$ the cubic equation \eqref{cubic}
has the following solution in the interval $\a_n \in [0,\pi]$
\begin{equation}
\cos\alpha_{n} \, = \, 0  \quad
\Rightarrow \quad \alpha_{n} \, = \, \frac{\pi}{2} \, .
\end{equation}
Expanding  \eqref{dimension-cartecian} around the unquenched result we have
\begin{equation}
\Delta \, = \,  \frac{1}{2} \, + \, \sqrt{\frac{9}{4} + l(l+2)} \, - \,
\frac{9}{512}\frac{l(l+2)}{\sqrt{\frac{9}{4} + l(l+2)}} \, \left( \frac{N_{f}}{k} \right)^{2}
\, + \, {\cal O}\left( \frac{N_{f}}{k} \right)^{3} \, .
\end{equation}
%


\subsection{Coupled modes}

In this subsection we will focus our attention on the fluctuations of the gauge and scalar fields,
which through their equations of motion appear to be coupled.
The equation of motion for the gauge field is given by
\be
\label{eq:flucA}
 {1 \ov \sqrt{\tilde{g}}} \, \partial_{\r}\left( \sqrt{\tilde{g}} \; \cG^{\m\r}\cG^{\n\s}f_{\m\n} \right)  \, + \,
 W \, \left( \, \partial_{r}\xi \, \delta_{0}^{\s} \, - \, \partial_{0}\xi \, \delta_{r}^{\s} \, \right) \, = \, 0 \, ,
\ee
while that for the scalar is
\be
\label{eq:flucxi}
{1 \ov \sqrt{\tilde{g}}} \, \partial_{\m}\left( \sqrt{\tilde{g}} \; \cG^{\m\n}\partial_{\n}\xi \right) \, - \,
\frac{2}{L^2}  \, {b^{2} \ov q} \, V \, \xi \, + \, {b^{2} \ov q}\,  {1 \ov L^{2}}\, W \, f_{0r} \, = \, 0 \, .
\ee
We consider the following ansatz for the fluctuations of the gauge fields and the scalar
(the only non-vanishing components of the gauge field are $\hat{A}_{r}$ and $\hat{A}_{i}$)
\begin{eqnarray} \label{coupled-ansatz}
&& \hat{A}_{r} \, = \, e^{\imath Et} \, \Om(\th_{2},\phi_{2},\psi) \, \phi(r) \, , \qquad
\hat{A}_{i} \, = \, e^{\imath Et} \, \partial_{i}\Om(\th_{2},\phi_{2},\psi) \, \tilde{\phi}(r)
\nonumber \\
&& \xi_{r} \, = \, e^{\imath Et} \, \Om(\th_{2},\phi_{2},\psi) \, z(r) \, .
\end{eqnarray}
Note that the fact that the vector index of $\hat A_i$ is due to the derivative on $\Om$, 
turns out after substitution into the equations of motion.
Then, from \eqref{eq:flucA} for $\s=0$ we obtain
\be
\label{eq:eq1}
 {b^{2} \ov q} {l(l+2) \ov \sin^{2}\a_{n}} \; \tilde{\phi}  \, = \,
 {r^{2} \ov P} \; \phi^{'}\, - \, \frac{\imath}{E} \, {L^{4}} \; P \; W \; r^{2} \; z^{'} \, ,
\ee
while if we set $\s=r$ in the same equation we have that
\be
\label{eq:eq2}
 {E^{2} \ov L^{4} P^{2}}\phi + {b^{2} \ov q} {l(l+2) \ov \sin^{2}\a_{n}} {r^{2} \ov L^{4}P}
 \left( \tilde{\phi}^{'} - \phi \right) \, - \,  \imath \, E \, W \, z \, = \, 0 \, .
\ee
Also from (\ref{eq:flucA}) for $\s=i$ we have
\be
 E^{2}\tilde{\phi} \, = \, - \, r^{2}\frac{d}{dr}\left[r^{2}\left(\tilde{\phi}^{'}-\phi\right)\right] \, ,
\ee
an equation that can be easily derived from \eqref{eq:eq1} and\eqref{eq:eq2} .
Using \eqref{eq:eq1} in order to
eliminate $\tilde{\phi}$ from (\ref{eq:eq2}), we have
\be
\label{eq:eq3}
 {d \ov dr}\left( r^{2} {d\phi \ov dr} \right) + {E^{2} \ov r^{2}}\phi - {b^{2} \ov q} {l(l+2) \ov \sin^{2}\a_{n}}P\phi - \imath {L^{4} \ov E} P^{2} W \left\{ {d \ov dr}\left( r^{2} {dz \ov dr} \right) + {E^{2} \ov r^{2}}z \right\}\, = \, 0 \, ,
\ee
while the equation of motion for the scalar, using \eqref{coupled-ansatz}, becomes
\be
\label{eq:eq4}
{d \ov dr}\left( r^{2} {dz \ov dr} \right) + {E^{2} \ov r^{2}}z - {b^{2} \ov q}
\left( 2V + {l(l+2) \ov \sin^{2}\a_{n}} \right) P \; z + \imath \frac{b^{2}}{q} E \; P \; W\phi \, = \, 0 \  .
\ee
At this point, we define the differential operator $\hat{\cO}$, which acts on functions as follows
\be
\label{eq:eq5}
 \hat{\cO}f={d \ov dr}\left( r^{2}{df \ov dr} \right) + {E^{2} \ov r^{2}}f
\ee
and make the following field redefinitions
\begin{equation}
\hat{z} \, = \,  \imath{L^{4} \ov E}P^{2} \; W \; z  \ , \qquad
\eta \, = \, \phi \, - \,  \imath \, {L^{4} \ov E} \, P^{2} \; W \; z \, = \,  \phi \, - \,  \hat{z} \, .
\end{equation}
Then \eqref{eq:eq3} and \eqref{eq:eq4} can be written in a more compact form as
\be
\label{eq:eq6}
 \left(\hat{\cO}\mathbb{I}_{2\times 2}-\cM\right)\begin{pmatrix}
                             \hat{z}\\
                             \eta
                           \end{pmatrix} = 0 \, ,
\ee
where the entries of the matrix $\cM$ are
\ba
 && \mathcal{M}_{11} \, = \, \frac{b^{2}}{q}\left( 2V + L^{4}W^{2}P^{2} +
 {l(l+2) \ov \sin^{2}\a_{n}} \right)P
 \nonumber\\ [7pt]
 && \mathcal{M}_{12} \, = \, \frac{b^{2}}{q}L^{4}W^{2}P^{3} \, , \qquad
  \mathcal{M}_{21} \, = \, \mathcal{M}_{22} \, = \, {b^{2} \ov q} {l(l+2) \ov \sin^{2}\a_{n}}P \, ,
\ea
while its eigenvalues are
\begin{equation}
\l_{\pm} \, = \, {P \ov 2} \, {b^{2} \ov q} \,  \Bigg[  2V + L^{4} \,P^{2} \, W^{2} + 2\, {l(l+2) \ov \sin^{2}\a_{n}}  \pm
 \sqrt{4  \, {l(l+2) \ov \sin^{2}\a_{n}} L^{4}P^{2}W^{2} + \left( 2V + L^{4}P^{2}W^{2} \right)^{2}} \Bigg] \, .
\end{equation}
Now, if $\psi_{\pm}$ are the eigenvectors of $\cM$ with eigenvalues $\l_{\pm}$, then \eqref{eq:eq6}
takes the form
\be
 {d \ov dr}\left( r^{2} {d\psi_{\pm} \ov dr} \right) + \left( {E^{2} \ov r^{2}} -\l_{\pm} \right)\psi_{\pm}\, = \, 0 \, ,
\ee
and in order to study the behavior of $\psi_{\pm}$ at large $r$ we assume that $\psi_{\pm} \sim r^{s}$.
Then from the above differential equation we find that $s$ should obey the following quadratic
equation
\be
 s(s+1) \, - \, \l_{\pm} \, = 0 \, ,
\ee
with solutions
\be
 s_{\pm} \, = \, {-1 \pm \sqrt{1+4\l_{\pm}} \ov 2} \, , \qquad s_{-}<s_{+} \, .
\ee
Noting that in the kinetic term in \eqref{2nd-Lang} for these type of fluctuations there is no extra overall factor of $r$,
we may safely use for  the conformal dimension the expression \cite{Witten:1998qj} with $d=1$ and $m^2=\l_+$, i.e.
\begin{equation} \label{dimension-coupled}
\D \, = \, {1 \ov 2} \, \left( \, 1\,  + \, \sqrt{ \, 1 \, + \,  4 \, \l_+} \, \right) \, .
\ee
In general, it is not a rational number and depends on the filling fraction $\nu$.
Notice, that unlike \eqref{dimension-cartecian}, for $l=0$ the conformal dimension $\Delta_{+}$ does depend on
the flavor number. As also argued in the appendix C,
for angular dependence of the form $\Om(\th_1,\phi_1)$ we may use the previous results by
simply performing the replacement \eqref{relllg}.
In the half filling fraction case  $\nu \, = \, \frac{1}{2}$
and expanding  \eqref{dimension-coupled} around the unquenched result we obtain that
\begin{eqnarray}
\Delta =  3 \, + \, l \, - \,  \frac{9}{512} \; \frac{(l+3)(l+2)(2l-1)}{(l+1)(5+2l)} \;
\left( \frac{N_{f}}{k} \right)^{2} \, + \,  {\cal O} \left( \frac{N_{f}}{k} \right)^{3}
\end{eqnarray}


\section{Conclusions and future directions} \label{sec:Conclusions}

In this paper we studied localized fermionic impurities to the unquenched ABJM
Chern--Simons-matter theory, which are realized through the addition of fields transforming in the
fundamental representations $(N,1)$ and $(1,N)$ of the $U(N) \times U(N)$ gauge group.
In the holographic approach the impurities are added by introducing probe D6-branes,
extending along the holographic coordinate and wrapping a five-dimensional submanifold inside a
squashed $\mathbb{CP}^3$ at constant values of the Minkowski directions.
The background RR field induces an electric gauge-field on the world-volume of the probe branes,
giving rise to a bundle of strings that form a flux tube which prevents the collapse of the wrapped brane.

We concentrated on the simplest of the configurations in which the flux tube starts from the boundary
of the  $AdS_4$ and reach the origin of the holographic coordinate. More general solutions
including the baryon vertex of the unquenched ABJM theory and hanging flux tubes, namely configurations
that reach a minimal value in the holographic coordinate and return to the boundary,
are left as open problems for future work.

The natural step forward was the investigation of the stability for the probe D6-branes, that introduce
the holographic impurities. We presented an analytic study for the fluctuations of those probes
in the unquenched ABJM. The fluctuations are separated in two categories.
The first contains just the decoupled fluctuations of the Cartesian coordinates while the second
the coupled fluctuations of the angular embedding function and the world-volume gauge field.
The coupled modes were shown to decouple by appropriate field redefinitions.
In this way we were able to determine the spectrum of conformal dimensions of the dual operators in the defect theory.
The novel feature of our analysis is that using the unquenched ABJM background we obtained expressions
of the conformal dimension that explicitly depend on the number of fundamental flavors, thus
generalizing the previously obtained results for the ABJM background \cite{Benincasa:2011zu}.

There are many interesting questions that follow from the analysis we presented in the recent paper,
and we would like to pursue some of them in the near future. In the quenched ABJM background there is the
robust proposition of \cite{Drukker:2008zx} that D6-branes on $AdS_2 \times {\cal M}_5$, where 
${\cal M}_5$ is a five-dimensional submanifold of  ${\mathbb C}{\mathbb P}^3$, 
holographically parametrize the Wilson lines in the antisymmetric representation of the gauge group. 
In turn these are natural candidates for the construction
of a gravity dual for an ABJM theory with fermionic impurities. 
Contrary to the quenched case, there is no such proof for the
unquenched ABJM, though we note the invariance under $n\to N-n$ in \eqref{change}.
It would be very interesting to pursue this issue further.

The addition of fermionic impurities in the unquenched ABJM background at finite
temperature \cite{Jokela:2012dw} will create a much richer structure. The analysis of the thermodynamic
properties, of both straight and hanging flux tubes, is expected to unveil a competition between the two
configurations. This in turn will lead to the existence of a dimerization
transition similar to the one presented in  \cite{Kachru:2009xf,Benincasa:2011zu}, but
now in a background that will include the non-trivial presence of fundamental flavors.


\section*{Acknowldegments}

We are grateful to A. Ramallo for sharing with us his unpublished notes and for valuable comments.
The research of G. Itsios has been co-financed by the ESF
and Greek national funds through the Operational Program ``Education and Lifelong Learning'' of the NSRF - Research Funding Program: ``Heracleitus II.
 Investing in knowledge in society through
the European Social Fund''. He would also like to thank the CFP of University of Porto
for hospitality within the framework of the LLP/Erasmus Placements 2011-2012.
D.~Z.~is funded by the FCT fellowship SFRH/BPD/62888/2009.
Centro de F\'{i}sica do Porto is partially funded by FCT through the projects
PTDC/FIS/099293/2008 and CERN/FP/116358/2010.
This research is implemented under the ``ARISTEIA" action of the  ``operational programme education and lifelong learning'' and is co-funded by the European Social Fund (ESF) and National Resources.


\appendix


\section{Fluctuation analysis}

\label{ap:apA}

In this appendix we will analyze the small perturbations around the flux tube configurations,
derived in section \ref{sec:FTC}. We will analytically obtain the second order lagrangian for
those fluctuations, which is the starting point of section \ref{sec:Fluct}.

We perturb a D6-brane probe as in \eqref{perturbation} and expand the DBI+WZ action
to second order  in the perturbations $\xi$, $f$ and $\chi$. Starting with the components
of the perturbed induced metric we write
\begin{equation}
g \, = \, \bar{g} \, + \, \hat{g} \, ,
\end{equation}
where $\bar{g}$ is the zeroth order induced metric and the perturbation $\hat{g}$ has the form
\begin{equation}
\hat{g}_{\m\n} \, = \, L^{2} \Big[ \,  r^{2} \, \partial_{\m}\chi \, \partial_{\n}\chi \, + \,
{q \ov b^{2}} \, \partial_{\m}\xi \, \partial_{\n}\xi  \, +  \, \xi \, \hat{g}_{\m\n}^{(1)}  \, +  \,
\xi^{2} \, \hat{g}_{\m\n}^{(2)}  \, \Big] + \dots  \, ,
\end{equation}
where the $\hat g_{\m\n}^{(1)}$ and  $\hat g_{\m\n}^{(2)}$ are given by (their indices
take values only in the angular part)
\begin{equation}
\begin{array}{l}
\begin{aligned}
\hat{g}_{ij}^{(1)} \, d\g^{i} \, d\g^{j} \, =  \, {q \ov 4b^{2}} \, \sin 2\a_{n} \,
\left( \, \om_{1}^{2} \, + \, \om_{2}^{2} \, + \, \om_{3}^{2} \, \right) \, +
{2 \ov b^{2}}  \left. \left[ \, \displaystyle{ E_{1} {\partial E_{1} \ov \partial\a}  \, + \,
E_{2} {\partial E_{2} \ov \partial\a} } \right] \right|_{\a=\a_{n}}
\end{aligned}\ ,
\\ [20pt]
\begin{aligned}
\hat{g}_{ij}^{(2)} \, d\g^{i} \, d\g^{j} & \, = \, {q \ov 4b^{2}} \, \cos 2\a_{n} \,
\left(\, \om_{1}^{2} \, + \, \om_{2}^{2} \, + \, \om_{3}^{2} \, \right)
\\
&\, +\,  {1 \ov b^{2}} \left. \left[ \, \left( {\partial E_{1} \ov \partial \a} \right)^{2}  \, +  \,
\left( {\partial E_{2} \ov \partial \a} \right)^{2} \, + \,
E_{1}{\partial^{2} \, E_{1} \ov \partial \a^{2}}  \, +  \,
E_{2}{\partial^{2} \, E_{2} \ov \partial \a^{2}}   \right] \right|_{\a=\a_{n}} \quad .
\end{aligned}
\end{array}
\end{equation}
The determinant of the DBI part  can be written as
\begin{equation}
\det (\,g \, + \, F \, ) \, = \, \det (\, \bar{g} \, + \, \bar{f} \, ) \, \det (\, \mathbbm{1} \, + \, X \, )
\ ,\qq  X \, = \, (\, \bar{g} \, + \, \bar{f}\, )^{-1} \, (\, \hat{g}\, + \, f \, ) \, .
\end{equation}
Hence the important step is to compute the components of matrix $X$  in the expansion
\begin{equation}
\sqrt{\det (\mathbbm{1} \, + \, X)} \, = \,
1 \, + \, {1 \ov 2} \, {\rm Tr}X \, -{1 \ov 4} \,  {\rm Tr}X^{2} \, + \, {1 \ov 8} \, ( {\rm Tr}X)^{2} \, + \, {\cal O}(X^{3}) \, .
\end{equation}
The matrix $(\, \bar{g} \, + \, \bar{f} \, )^{-1}$ can be written in a block diagonal form
\begin{equation}
(\, \bar{g} \, + \, \bar{f} \, )^{-1} \, = \,
\left(
\begin{array}{c|c}
\left. \cG^{-1} \, + \, \cJ \right|_{0r} & 0\\ \hline
         0                                        & \cG^{ij}
\end{array}
\right) \, ,
\end{equation}
where $\cG^{-1}$ and $\cJ$ are its symmetric and antisymmetric parts, respectively.
The non-zero elements of those matrices are
\begin{equation} \label{sym}
\begin{array}{lll}
\begin{aligned}
\cG^{00} \, = \, - \, \frac{\sin^{6}\a_{n} \, + \, C_{n}(\a_{n})^{2}}{r^{2} \, L^{2} \, \sin^{6}\a_{n}} \, , &
\quad \cG^{rr} \, = \, \frac{\sin^{6}\a_{n} \, + \, C_{n}(\a_{n})^{2}}{L^{2} \, \sin^{6}\a_{n}} \, r^{2} \, , &
\quad \cG^{ij} \, = \, L^{-2} \, \tilde{g}^{ij} \,
\end{aligned}
\end{array}
\end{equation}
and
\begin{equation} \label{antisym}
\cJ^{0r} \, = \, - \, \cJ^{r0} \, = \, \frac{C_{n}(\a_{n}) \, \sqrt{\sin^{6}\a_{n} \, + \,  C_{n}(\a_{n})^{2}}}
{L^{2} \, \sin^{6}\a_{n}} \, .
\end{equation}
The matrix elements that contribute to the ${\rm Tr} X$ are
\ba
X^{0}_{0} && \, = \, - \, {\sin^{6}\a_{n} \, + \, C_{n}(\a_{n})^{2} \ov r^{2} \, \sin^{6}\a_{n}}
\left\{ \, r^{2} \, (\partial_{0}\chi)^{2} \, + \, {q \ov b^{2}} \, (\partial_{0}\xi)^{2} \, \right\} \ ,
\nonumber\\[0.3cm]
 && \, + \,  {C_{n}(\a_{n}) \, \sqrt{\sin^{6}\a_{n} \, + \, C_{n}(\a_{n})^{2}} \ov \sin^{6}\a_{n}}
 \left\{ \, r^{2} \, \partial_{0} \chi \, \partial_{r}\chi \, + \, {q \ov b^{2}} \, \partial_{0}\xi \, \partial_{r}\xi \,
 - \,  L^{-2} \, f_{0r} \, \right\}\ ,
\nonumber\\[0.3cm]
X^{r}_{r} && \, = \, {\sin^{6}\a_{n} \, + \, C_{n}(\a_{n})^{2} \ov \sin^{6}\a_{n}} \,  r^{2}
\left\{ \, r^{2}(\partial_{r}\chi)^{2} \, + \, {q \ov b^{2}} \, (\partial_{r}\xi)^{2} \, \right\}
\\[0.3cm]
 && \, - \, {C_{n}(\a_{n}) \,\sqrt{\sin^{6}\a_{n} \, + \, C_{n}(\a_{n})^{2}} \ov \sin^{6}\a_{n}}
\left\{ \, r^{2} \, \partial_{0}\chi \, \partial_{r}\chi \, + \, {q \ov b^{2}} \, \partial_{0}\xi \, \partial_{r}\xi  \, +
\,  L^{-2} \, f_{0r} \, \right\}\ ,
\nonumber\\[0.3cm]
X^{i}_{j} && \, = \, r^{2} \, \tilde{g}^{ik} \, \partial_{k}\chi \, \partial_{j}\chi \, + \,  {q \ov b^{2}} \,
\tilde{g}^{ik} \, \partial_{k}\xi \, \partial_{j}\xi \, + \,  \xi \, \left( M^{(1)} \right)^{i}_{j}  \, + \,
\xi^{2} \, \left( M^{(2)} \right)^{i}_{j} \, + \,  L^{-2} \, \tilde{g}^{ik} \, f_{kj} \, ,\nonumber
\ea
where we have defined the matrices
\begin{equation}
\left( \, M^{(i)} \, \right)^{i}_{j} \, = \, \tilde{g}^{ik} \, \hat{g}^{(i)}_{kj}\ , \qquad i\, = \, 1,2 \, .
\end{equation}
The metric $\tilde{g}_{ij}$  corresponds to the five-dimensional space whose line element is given by
\eqref{M5}.
In order to calculate the trace of $X^2$, we need to compute the non-diagonal elements of $X$ up to first order in the fluctuations.
The matrix elements that contribute to ${\rm Tr} X^2$ are
\ba
 && X^{0}_{r} \, = \, - \, {\displaystyle{ \sin^{6}\a_{n} \, + \, C_{n}(\a_{n})^{2} \ov r^{2} \, L^{2} \, \sin^{6}\a_{n} }}
\, f_{0r}
\nonumber\\[0.3cm]
 && X^{r}_{0} \, = \, - \, {\displaystyle{ \sin^{6}\a_{n} \, \, + \, C_{n}(\a_{n})^{2} \ov L^{2} \, \sin^{6}\a_{n} }} \,
r^{2} \, f_{0r}
\nonumber\\[0.3cm]
 && X^{0}_{i} \, = \, - \, {\displaystyle{ \sin^{6}\a_{n} \, + \, C_{n}(\a_{n})^{2} \ov r^{2} \, L^{2} \, \sin^{6}\a_{n} }} \,
 f_{0i} \, + \, {\displaystyle{ C_{n}(\a_{n}) \, \sqrt{\sin^{6}\a_{n} \, + \, C_{n}(\a_{n})^{2}} \ov
 L^{2} \, \sin^{6}\a_{n} }} \, f_{ri}
\\[0.3cm]
 && X^{i}_{0} \, = \, L^{-2} \, \tilde{g}^{ij} \, f_{j0}
\nonumber\\[0.3cm]
 && X^{r}_{i} \, = \, {\displaystyle{ \sin^{6}\a_{n} \, + \, C_{n}(\a_{n})^{2} \ov L^{2} \, \sin^{6}\a_{n} }} \, r^{2}\,
f_{ri}  \, - \,  {\displaystyle{ C_{n}(\a_{n}) \, \sqrt{\sin^{6}\a_{n} \, + \, C_{n}(\a_{n})^{2}}
\ov L^{2} \, \sin^{6}\a_{n} }} \, f_{0i}
\nonumber\\[0.3cm]
 && X^{i}_{r} \, = \, L^{-2} \, \tilde{g}^{ij} \, f_{jr}
\nonumber
\ea
Putting everything together we find that the $TrX$ is given by
\begin{eqnarray}
{\rm  Tr} X &=& \, - \,  2\frac{C_{n}(\a_{n}) \, \sqrt{\sin^{6}\a_{n} \, + \, C_{n}(\a_{n})^{2}}}{\sin^{6}\a_{n}} \,
 L^{-2} f_{0r} \, + \,  6\cot\a_{n} \, \xi  \, + \,  \frac{1 \, + \, (3q \, - \,1)
 \cos 2\a_{n}}{q \, \sin^{2}\a_{n}} \, \xi^{2}
 \nonumber\\[0.3cm]
 &+& \, L^{2} \, r^{2} \, \cG^{\m\n} \, \partial_{\m} \chi \, \partial_{\n}\chi \, + \,
 \frac{q}{b^{2}} \, L^{2} \, \cG^{\m\n} \, \partial_{\m}\xi \, \partial_{\nu}\xi \, ,
\end{eqnarray}
where we have used the following equations
\begin{equation}
{\rm Tr} M^{(1)} \, = \, 6\cot\a_{n} \ ,\qq
 {\rm Tr} M^{(2)} \, = \, \frac{1+(3q-1)\cos 2\a_{n}}{q \, \sin^{2}\a_{n}} \, .
\end{equation}
The ${\rm Tr}(X^{2})$ is given by
\begin{eqnarray}
Tr(X^{2}) &=& 2L^{-4} \, \frac{\sin^{6}\a_{n}+2C_{n}(\a_{n})^{2}}{\sin^{6}\a_{n}} \; \frac{\sin^{6}\a_{n} \, + \,
C_{n}(\a_{n})^{2}}{\sin^{6}\a_{n}}f_{0r}^{2} + 2\xi^{2} \frac{1+3q+(3q-1)\cos 2\a_{n}}{q\sin^{2}\a_{n}} \nonumber\\[0.3cm]
&+& L^{-4} \left\{ 2\; \frac{\sin^{6}\a_{n}+C_{n}(\a_{n})^{2}}{\sin^{6}\a_{n}}\frac{f^{2}_{0i}}{r^{2}} - 2\; \frac{\sin^{6}\a_{n}+C_{n}(\a_{n})^{2}}{\sin^{6}\a_{n}}r^{2}f^{2}_{ri} - f^{2}_{kj} \right\} \, ,
\end{eqnarray}
while for the $({\rm Tr} X)^{2}$ we have
\begin{eqnarray}
({\rm Tr} X)^{2} &=& 4L^{-4}  \, \frac{ \sin^{6}\a_{n} \, + \, C_{n}(\a_{n})^{2}}{\sin^{12}\a_{n}}\,
C_{n}(\a_{n})^2\,  f^{2}_{0r} \, + \,  36\cot^{2}\a_{n} \, \xi^{2}
\nonumber\\[0.3cm]
&-& 24L^{-2} \,  \frac{C_{n}(\a_{n})\, \sqrt{\sin^{6}\a_{n} \, + \, C_{n}(\a_{n})^{2}}}{\sin^{6}\a_{n}} \,
\cot\a_{n} \; \xi \; f_{0r} \, .
\end{eqnarray}
Putting everything together we calculate the expression of $\sqrt{\det (\mathbbm{1}+X)}$
\begin{eqnarray} \label{DBI-det}
\sqrt{\det (\mathbbm{1}+X)} &=& 1 \, - \,  \frac{C_{n}(\a_{n})\sqrt{\sin^{6}\a_{n}+C_{n}(\a_{n})^{2}}}
{\sin^{6}\a_{n}}L^{-2}f_{0r} + 3\cot\a_{n} \; \xi + \frac{1}{2} \, L^{2}r^{2}\cG^{\m\n}
\partial_{\m}\chi\partial_{\nu}\chi
\nonumber\\[0.3cm]
&+& \frac{1}{2} \, \frac{q}{b^{2}} \, L^{2}\cG^{\m\n}\partial_{\m}\xi\partial_{\n}\xi \, +  \,
\frac{3}{2} \, (2\cot^{2}\a_{n}-1)\xi^{2} \, + \,  \frac{1}{4} \, \cG^{\m\r}\cG^{\n\s}f_{\m\n}f_{\r\s}
\\[0.3cm]
&-& 3L^{-2} \, \frac{C_{n}(\a_{n})\sqrt{\sin^{6}\a_{n}+C_{n}(\a_{n})^{2}}}{\sin^{6}\a_{n}} \,
\cot\a_{n} \; \xi \; f_{0r} +\dots  .
\nonumber
\end{eqnarray}
Since
\begin{equation}
\det (\bar{g} \, + \, \bar{f}) \, =  \, - \, {L^{14} \, \sin^{6}\a_{n} \ov \sin^{6}\a_{n} \, + \, C_{n}(\a_{n})^{2}}
\,  \cdot  \, \det\tilde{g} \, ,
\end{equation}
the DBI part of the Lagrangian density is
\begin{equation} \label{ap:DBI}
\cL_{DBI} \, = \, - \, T_{6} \; {\pi^{2} \, N \, b^{6} \, L^{2} \ov q^{2}} \; {\sin^{3}\a_{n} \ov
\sqrt{\sin^{6}\a_{n} \, + \, C_{n}(\a_{n})^{2}}} \, \sqrt{\tilde{g}} \, \sqrt{\det (\mathbbm{1}+X)} \, ,
\end{equation}
where ${\tilde{g}}$ is the determinant of \eqref{M5}
\begin{equation}
\sqrt{\tilde{g}} \, = \, {q^{3/2} \ov 8\,  b^{5}} \, \sin^{3}\a_{n} \, \sin\th_{1} \, \sin\th_{2} \, .
\end{equation}
What remains is the computation of the WZ part. Using the conventions of section \ref{density} we have
\begin{equation}
 \cL_{WZ} \, = \, - \, T_{6} \; {\pi^{2}N \ov 8} \; \sin\th_{1} \, \sin\th_{2} \; C(\a) \; F_{0r} \, ,
\end{equation}
where $F_{0r}=\bar{f}_{0r}+f_{0r}$ and the function $C(\a)$ has to be expanded around $\a_{n}$
\begin{equation}
C(\a) \, = \,  C(\a_{n}) \, - \,  3\sin^{3}\a_{n} \; \xi \, - \,
\frac{9}{2} \, \sin^{2}\a_{n} \, \cos\a_{n} \; \xi^{2} + \dots  \, .
\end{equation}
Putting everything together, the WZ part becomes
\begin{eqnarray} \label{ap:WZ}
\cL_{WZ} &=& - \,T_{6} \; {\pi^{2}N \ov 8} \; \sin\th_{1}\sin\th_{2} \left\{ {L^{2}C_{n}(\a_{n}) \ov
\sqrt{\sin^{6}\a_{n}+C_{n}(\a_{n})^{2}}} \, C(\a_{n}) \, + \,  C(\a_{n}) \, f_{0r} \right.
\\[0.3cm]
&&\left. - \, {3L^{2}\sin^{3}\a_{n} \, C_{n}(\a_{n}) \ov \sqrt{\sin^{6}\a_{n}+C_{n}(\a_{n})^{2}}} \; \xi \,  - \,
3\sin^{3}\a_{n} \; f_{0r} \; \xi  \, - \,  {9 \ov 2}L^{2} \, {\sin^{2}\a_{n} \, \cos\a_{n} \, C_{n}(\a_{n}) \ov
\sqrt{\sin^{6}\a_{n}+C_{n}(\a_{n})^{2}}}
\; \xi^{2} \right\} + \dots \, .
\nonumber
\end{eqnarray}
Finally, summing the DBI and WZ parts we obtain the result
\ba
\mathcal{L}_{DBI} + \mathcal{L}_{WZ} &=& - T_{6}  {\pi^{2}  N  b^{6}  L^{2} \ov q^{2}}  {\sin^{3}\a_{n} \ov \sqrt{\sin^{6}\a_{n} + C_{n}(\a_{n})} } \sqrt{\tilde{g}}
\left\{ 1 + \frac{\sqrt{q}}{b}\frac{C_{n}(\a_{n})}{\sin^{6}\a_{n}}C(\a_{n}) \right.
\nonumber\\[0.3cm]
&+& \left. \left(\frac{\sqrt{q}}{b} C(\a_{n}) -  C_{n}(\a_{n})\right)\frac{\sqrt{\sin^{6}\a_{n}+C_{n}(\a_{n})^{2}}}
{\sin^{6}\a_{n}} L^{-2}f_{0r}  \right.
\label{flloicr}
\\[0.3cm]
&+& \left. 3\left(\cot\a_{n} - \frac{\sqrt{q}}{b}\frac{C_{n}(\a_{n})}{\sin^{3}\a_{n}} \right) \xi + \frac{1}{2}  L^{2}r^{2}\cG^{\m\n} \partial_{\m}\chi\partial_{\nu}\chi
+ \frac{1}{2}  \frac{q}{b^{2}}  L^{2}\cG^{\m\n}\partial_{\m}\xi\partial_{\n}\xi \right.
\nonumber\\[0.3cm]
&+& \left. \left( \frac{3}{2}  (2\cot^{2}\a_{n}-1) -\frac{9}{2}\frac{\sqrt{q}}{b}\frac{\cos\a_{n}}{\sin^{4}\a_{n}}C_{n}(\a_{n}) \right) \xi^{2}
+ \frac{1}{4}  \cG^{\m\r}\cG^{\n\s}f_{\m\n}f_{\r\s} \right.
\nonumber\\[0.3cm]
&-& \left. 3L^{-2} {\sqrt{\sin^{6}\a_{n} + C_{n}(\a_{n})} \ov \sin^{6}\a_{n}} \left( {\sqrt{q} \ov b}\sin^{3}\a_{n} + C_{n}(\a_{n})\cot\a_{n} \right) \xi \, f_{0r} \right\} .
\nonumber
\ea
In the above action the first two lines can be dropped containing either constant or linear or total derivative terms.
The term linear in $\xi$ also vanishes upon using the condition \eqref{lnna}. The remaining terms constitute the
action \eqref{2nd-Lang} used in the main text.


\section{Solution of the differential equation for $R(r)$}

\label{ap:apB}

The differential equation (\ref{eq:ODE}) is of the form
\be
{d \ov dr} \left( r^{4} {dR \ov dr} \right) + (E^{2} -A r^{2})R = 0 \, ,
\label{dgflr4}
\ee
where $A$ is a constant.
To solve this differential equation we first study the asymptotic behavior of $R(r)$ at large $r$.
Setting $R(r) \sim r^{\l}$ we end up with the condition
\begin{equation}
\l(\l+3) \, = \, A \qq \Longrightarrow \qq \l_{\pm} \, = \,  - \, {3 \ov 2} \pm \sqrt{\, {9 \ov 4} \, + \,  A} \, .
\end{equation}
Now that we know the asymptotic behavior of $R(r)$, we make the ansatz $R(r)=r^{\l}f({1 \ov r})$ and
obtain the following differential equation for the function $f=f(u)=f(1/r)$
\begin{equation}
u^{2} \, {d^{2}f \ov du^{2}} \, - \,  2(\l+1) \, u \, {df \ov du} \, + \,  E^{2}  \, u^{2} \, f \,  = \,  0 \, .
\ee
Now we observe that if we substitute $f(u)=u^{\l + {3 \ov 2}}g(u)$ the above equation becomes
\be
u^{2} \, {d^{2}g \ov du^{2}} \, + \,  u{dg \ov du}  \, + \,  (E^2 u^2 \, - \, t^{2}) \, g \, = \,  0 \, ,
\ee
where $t^{2} = (\l + {3 \ov 2})^{2} = {9 \ov 4} + A$ , which is the Bessel differential
equation with solution
\be
g(u) = C_{1}J_{\sqrt{ {9 \ov 4} + A}}(Eu) + C_{2}Y_{\sqrt{ {9 \ov 4} + A}}(Eu) \, .
\ee
Finally, the solution for $R(r)$ is
\be
R(r)=r^{-3/2} \left( C_{1}J_{\sqrt{ {9 \ov 4} + A}} \left( {E \ov r} \right) +
C_{2}Y_{\sqrt{ {9 \ov 4} + A}} \left( {E \ov r} \right) \right) \, ,
\ee
which has the correct asymptotic behavior at large $r$
\be
R(r)=C_{1}r^{-{3 \ov 2} + \sqrt{ {9 \ov 4} + A}} + C_{2}r^{-{3 \ov 2} - \sqrt{ {9 \ov 4} + A}} \, .
\ee
In our case the constant $A$ appearing in \eqref{dgflr4} is
\be
 A = {b^{2} \ov q} \; {l(l+2) \ov \sin^{2}\a_{n}} \; P \ .
\ee


\section{The Laplacian on the five dimensional manifold $M_{5}$}

\label{ap:apC}

The action of the Laplacian of the five-dimensional manifold $M_{5}$,
with metric $\tilde{g}_{ij}$ \eqref{M5} on a scalar depending on all the coordinates of $M_{5}$ is
\ba \label{M5-Laplacian}
 \nabla^{2}_{M_{5}}f &&= \frac{c_{1}}{4\sin^{2}\frac{\alpha_{n}}{2}}\nabla^{2}_{S^{3}}f   +  \frac{c_{1}c_{2}}{2}\frac{1}{\sin\theta_{1}}\partial_{\theta_{1}}\left( \sin\theta_{1}\partial_{\theta_{1}}f \right)  +   \frac{c_{1}c_{2}}{2}\frac{1}{\sin^{2}\theta_{1}}\partial^{2}_{\phi_{1}}f   \nonumber\\[0.2cm]
                     &&- \frac{c_{1}\sin(\phi_{1}-\psi)}{\sin\theta_{1}}\partial_{\theta_{1}}\left( \sin\theta_{1}\partial_{\theta_{2}}f \right)  -   \frac{c_{1}\sin(\phi_{1}-\psi)}{\sin\theta_{2}}\partial_{\theta_{2}}\left( \sin\theta_{2}\partial_{\theta_{1}}f \right) \nonumber\\[0.2cm]
                     &&- \frac{c_{1}\cos(\phi_{1}-\psi)}{\sin\theta_{1}\sin\theta_{2}}\partial_{\theta_{1}}\left( \sin\theta_{1}\partial_{\phi_{2}}f \right)   -   \frac{c_{1}\cos(\phi_{1}-\psi)}{\sin\theta_{2}}\partial_{\phi_{2}}\partial_{\theta_{1}}f \\[0.2cm]
                     &&+ \frac{c_{1}\cos(\phi_{1}-\psi)\cot\theta_{2}}{\sin\theta_{1}}\partial_{\theta_{1}}\left( \sin\theta_{1}\partial_{\psi}f \right)  +   c_{1}\cot\theta_{2}\partial_{\psi}\left( \cos(\phi_{1}-\psi)\partial_{\theta_{1}}f \right) \nonumber\\[0.2cm]
                     &&- c_{1}\cot\theta_{1}\partial_{\phi_{1}}\left( \cos(\phi_{1}-\psi)\partial_{\theta_{2}}f \right)    -    \frac{c_{1}\cos(\phi_{1}-\psi)\cot\theta_{1}}{\sin\theta_{2}}\partial_{\theta_{2}}\left( \sin\theta_{2}\partial_{\phi_{1}}f \right) \nonumber\\[0.2cm]
                     &&+ c_{1}\frac{\cot\theta_{1}}{\sin\theta_{2}}\partial_{\phi_{1}}\left( \sin(\phi_{1}-\psi)\partial_{\phi_{2}}f \right)   +    c_{1}\frac{\cot\theta_{1}\sin(\phi_{1}-\psi)}{\sin\theta_{2}}\partial_{\phi_{2}}\partial_{\phi_{1}}f  \nonumber\\[0.2cm]
                     &&+ \frac{c_{1}}{\sin\theta_{1}\sin\theta_{2}}\partial_{\phi_{1}}\left( \left( \sin\theta_{1}\sin\theta_{2}-\cos\theta_{1}\cos\theta_{2}\sin(\phi_{1}-\psi) \right)\partial_{\psi}f \right) \nonumber\\[0.2cm]
                     &&+ \frac{c_{1}}{\sin\theta_{1}\sin\theta_{2}}\partial_{\psi}\left( \left( \sin\theta_{1}\sin\theta_{2}-\cos\theta_{1}\cos\theta_{2}\sin(\phi_{1}-\psi) \right) \partial_{\phi_{1}}f \right) \nonumber
\ea
where the last term is the Laplacian of the 3-sphere
\begin{equation}
\nabla^{2}_{S^{3}}f \, = \, 4 \, \left\{ \frac{1}{\sin\th_{2}} \, \partial_{\th_{2}}\left( \, \sin\th_{2} \,
\partial_{\th_{2}}f \, \right)  \, + \, \left( \, \frac{1}{\sin\th_{2}} \, \partial_{\phi_{2}} \, - \,
\cot\th_{2} \, \partial_{\psi} \, \right)^{2}f \, + \, \partial_{\psi}^{2}f \, \right\} \, ,
\end{equation}
and the constants $c_{1}, c_{2}$ are given by
\be
c_{1} \, = \, \frac{b^{2}}{q} \, \frac{1}{\cos^{2}\frac{\a_{n}}{2}} \ ,\qquad
c_{2} \, = \, 1 \, + \, q \, + \, (q \, - \, 1) \, \cos \a_{n} \, .
\ee
Solving the corresponding eigenvalue problem in full generality seems very difficult.
Instead, we look for configurations of the function $f$, in which the Laplacian acts on, that do not
depend on all variables. We found two consistent truncations that simplify our eigenvalue problem and we present
them in the following subsections.


\subsection{Solutions of the form $f=\Om(\theta_{2},\phi_{2},\psi)$}

\label{ap:apC1}

In the analysis that we presented in the previous sections of this paper we focused on configurations
of the form  $f=\Om(\theta_{2},\phi_{2},\psi)$. For such configurations the above expression for the Laplacian becomes
\begin{equation}
\begin{aligned}
  \nabla^{2}_{M_{5}}\Om &=\frac{c_{1}}{4\sin^{2}\frac{\alpha_{n}}{2}}\nabla^{2}_{S^{3}}\Om   -   \frac{c_{1}\sin(\phi_{1}-\psi)}{\sin\theta_{1}}\partial_{\theta_{1}}\left( \sin\theta_{1}\partial_{\theta_{2}}\Om \right)\\[0.2cm]
                        & -\frac{c_{1}\cos(\phi_{1}-\psi)}{\sin\theta_{1}\sin\theta_{2}}\partial_{\theta_{1}}\left( \sin\theta_{1}\partial_{\phi_{2}}\Om \right)   +   \frac{c_{1}\cos(\phi_{1}-\psi)\cot\theta_{2}}{\sin\theta_{1}}\partial_{\theta_{1}}\left( \sin\theta_{1}\partial_{\psi}\Om \right) \\[0.2cm]
                        &-c_{1}\cot\theta_{1}\partial_{\phi_{1}}\left( \cos(\phi_{1}-\psi)\partial_{\theta_{2}}\Om \right)   +   c_{1}\frac{\cot\theta_{1}}{\sin\theta_{2}}\partial_{\phi_{1}}\left( \sin(\phi_{1}-\psi)\partial_{\phi_{2}}\Om \right) \\[0.2cm]
                        & +\frac{c_{1}}{\sin\theta_{1}\sin\theta_{2}}\partial_{\phi_{1}}\left( \left( \sin\theta_{1}\sin\theta_{2}-\cos\theta_{1}\cos\theta_{2}\sin(\phi_{1}-\psi) \right)\partial_{\psi}\Om \right) \, .
\end{aligned}
\end{equation}
Notice that there are still derivatives to be taken with respect to the angles $\th_1$ and $\phi_1$ as well as explicit
dependence on these angles. Consistency of our truncation requires that all such dependencies drop out completely which indeed
turns out to be the case as we are left with
\begin{equation}
\label{3Laplacian}
\nabla_{M_{5}}^{2}\Om \, = \, \frac{c_{1}}{4\sin^{2}\frac{\a_{n}}{2}} \, \nabla^{2}_{S^{3}}\Om \, .
\end{equation}
In this way our eigenvalue problem is reduced to that of $\nabla^{2}_{S^{3}}$, a well known operator,
which has eigenvalues $\lambda=-l(l+2)$ where $l=0,1,2,\ldots$.


\subsection{Solutions of the form $f=\Om(\theta_{1},\phi_{1})$}

\label{ap:apC2}

Another truncation that simplifies the eigenvalue problem is an ansatz of the form $f=\Om(\th_{1},\phi_{1})$.
Then \eqref{M5-Laplacian} simplifies as
\ba
  \nabla_{M_{5}}^{2}\Om &&= \frac{c_{1}c_{2}}{2}\frac{1}{\sin\theta_{1}}\partial_{\theta_{1}}\left( \sin\theta_{1}\partial_{\theta_{1}}\Om \right) + \frac{c_{1}c_{2}}{2}\frac{1}{\sin^{2}\theta_{1}}\partial^{2}_{\phi_{1}}\Om -\frac{c_{1}\sin(\phi_{1}-\psi)}{\sin\theta_{2}}\partial_{\theta_{2}}\left( \sin\theta_{2}\partial_{\theta_{1}}\Om \right) \nonumber\\[0.2cm]
                        &&- \frac{c_{1}\cos(\phi_{1}-\psi)\cot\theta_{1}}{\sin\theta_{2}}\partial_{\theta_{2}}\left( \sin\theta_{2}\partial_{\phi_{1}}\Om\right) + c_{1}\cot\theta_{2}\partial_{\psi}\left( \cos(\phi_{1}-\psi)\partial_{\theta_{1}}\Om \right)\\[0.2cm]
                        &&+ \frac{c_{1}}{\sin\theta_{1}\sin\theta_{2}}\partial_{\psi}\left( \left( \sin\theta_{1}\sin\theta_{2}-\cos\theta_{1}\cos\theta_{2}\sin(\phi_{1}-\psi) \right) \partial_{\phi_{1}}\Om \right) \, . \nonumber
\ea
Again all dependence on $\th_1,\phi_1$ and $\psi$, finally obtaining that
\be
\nabla_{M_{5}}^{2}\Om = \frac{c_{1}c_{2}}{2}\frac{1}{\sin\theta_{1}}\partial_{\theta_{1}}\left( \sin\theta_{1}\partial_{\theta_{1}}\Om \right) + \frac{c_{1}c_{2}}{2}\frac{1}{\sin^{2}\theta_{1}}\partial^{2}_{\phi_{1}}\Om = \frac{c_{1}c_{2}}{2}\nabla_{S^{2}}^{2}\Om \, .
\label{nab52}
\ee
We see that configurations of the form $f=f(\th_{1},\phi_{1})$ lead to the well known eigenvalue problem of
the Laplace operator on the unit $S^2$, with  eigenvalues $\lambda=-l(l+1)$,
where $l=0,1,2,\ldots$

The results we have obtained for the fluctuations using the ansatz $f=\Om(\th_2,\phi_2,\psi)$ can be
trivially extended to the case when $f=\Om(\th_1,\phi_1)$. We simply have to compare \eqref{3Laplacian} and \eqref{nab52} and
make the replacement
\be
l(l+2)\to 4 \sin^2{\a_n\ov 2} \left(  \sin^2{\a_n\ov 2} + q  \cos^2{\a_n\ov 2}\right)l(l+1)\ .
\label{relllg}
\ee


\end{document}